# Faculty mobility reallocates research capacity within persistent institutional hierarchies

**Erjia Yan, Chaoqun Ni**

## Significance statement

Universities recruit faculty on the premise that better-resourced environments will amplify scientific output. Using a large-scale longitudinal analysis of faculty at U.S. research universities, we show that mobility instead operates primarily as a reallocative process. Prestigious universities disproportionately attract faculty, but movers' publication and citation trajectories largely continue along paths established before relocation. The clearest post-move change is in collaboration: faculty form new coauthor ties. These findings suggest that faculty mobility can intensify institutional inequality by shifting where research capacity is institutionally located, without broad productivity gains for individual movers.

## Abstract

Faculty mobility is often understood as a mechanism through which universities redistribute scientific talent and potentially improve research performance. Yet the system-level structure of mobility and its association with individual research trajectories have rarely been examined together. We link longitudinal faculty rosters from U.S. research universities to OpenAlex publication records and study 11,535 tenure-system faculty members who changed institutions between 2011 and 2020, with a comparison group of more than 200,000 non-moving faculty members. A directed network of faculty moves reveals a strongly hierarchical market: high-prestige institutions are net importers, lower-prestige institutions are net exporters, and the mobility hierarchy closely parallels the hierarchy observed in faculty hiring. However, event-study models that account for pre-move trajectories show little evidence of sustained post-move gains in publication volume, citation impact, or top-cited publication rates, including among upward moves to more prestigious institutions. The most consistent post-move change is collaborative: movers form new coauthor ties. We also observe modest increases in the share of papers with positive CD-index values. These patterns suggest that faculty mobility primarily reallocates existing research capacity within a persistent institutional hierarchy rather than systematically altering individual research trajectories.

## Introduction

Universities devote substantial resources to recruiting faculty, often on the assumption that institutional environments amplify scientific performance (1, 2). Better-resourced universities may provide stronger infrastructure, larger research groups, more visible intellectual communities, and access to collaborators and graduate students (3, 4). From this perspective, mobility is not only a labor-market mechanism but also a possible engine of scientific production: relocating faculty to stronger environments should increase the volume or impact of their research output (3, 5).

The causal effects of institutional mobility on research performance remain uncertain. Faculty often change institutions during periods of career transition (6, 7), making it difficult to distinguish the effect of relocation from the continuation of underlying career trajectories.

Apparent post-move gains may simply reflect preexisting momentum, while declines may reflect regression to the mean. Much of the existing literature therefore documents associations between mobility and performance without clearly isolating whether institutional environments themselves alter research trajectories (8, 9).

These competing interpretations map onto two broad views. The first emphasizes institutional amplification: moving to a stronger environment should increase output by improving access to resources, collaborators, and visibility (2, 4, 10). The second emphasizes portable human capital: productivity primarily reflects accumulated expertise, research agendas, and collaboration networks that faculty carry with them across institutions. Under this view, mobility reallocates existing research capacity more than it transforms it (11, 12). Distinguishing between these views is difficult because faculty do not move randomly. Prior studies have documented associations between mobility, prestige, and performance, but the field still lacks a system-wide account that connects two levels of the problem: how faculty mobility reorganizes the institutional hierarchy of science, and whether relocation is associated with departures from individual research trajectories.

To adjudicate between these views, we assemble a near-census panel of faculty mobility across U.S. research universities by linking the Academic Analytics Research Center faculty roster (13) to OpenAlex publication records (14). The resulting dataset includes 11,535 tenure-system faculty who changed institutions, 12,207 moves across more than 300 universities between 2011 and 2020, and a comparison group of more than 200,000 non-moving faculty. The panel spans five years before and five years after each move and includes more than 13 million publications.

This paper asks two questions. First, does faculty mobility reinforce institutional hierarchy by moving faculty toward institutions that are already prestigious? Second, are moves associated with discontinuous changes in publication, citation, and collaboration trajectories after accounting for pre-move momentum? We find that mobility is highly hierarchical: prestigious universities are net importers of faculty, less prestigious universities are net exporters, and the mobility-derived prestige hierarchy closely reproduces the hierarchy observed in faculty hiring. At the individual level, however, movers largely remain close to the publication and citation trajectories established before relocation. The clearest change is not in output but in collaboration, as movers form new coauthor ties after changing institutions. These patterns suggest that mobility contributes to cumulative advantage by concentrating faculty research capacity at high-prestige institutions while leaving individual publication and citation trajectories largely intact.

## Results

### Faculty mobility reveals a hierarchical prestige structure

We first characterize institutional mobility as a directed network in which nodes are universities and weighted edges are observed faculty transitions. We infer an institutional prestige hierarchy from observed faculty moves using PageRank, a centrality measure that assigns greater weight to institutions that receive faculty from other highly ranked institutions (15-19). Because this prestige measure is derived from mobility itself, we treat it as a structural representation of the mobility system rather than as an exogenous institutional quality measure. We therefore validate this measure against independent hiring-based rankings and assess robustness using alternative prestige metrics and null models.

The mobility network is strongly stratified by institutional prestige (Fig. 1). After partitioning institutions into PageRank quintiles, 44.8% of moves occur within the same prestige tier, indicating substantial circulation among peer institutions. Mobility is also directionally asymmetric: 35.2% of moves ascend the prestige hierarchy, whereas 20.0% descend (Fig. 1A). These patterns indicate that faculty moves are not evenly distributed across the institutional system but are organized by both assortative exchange and directional sorting toward higher-prestige destinations.

Faculty inflows are especially concentrated at the top (Fig. 1B). The highest prestige tier captures a disproportionate share of inflows while accounting for a smaller share of outflows, making it the only tier with a large positive net gain. Lower tiers are net exporters. The empirical transition structure also exceeds what would be expected from a degree-preserving null model that maintains overall origin and destination volumes (Fig. 1C): observed within-tier mobility is higher than the null expectation (44.8% vs. 38.4%), and the observed upward-minus-downward asymmetry is larger (0.152 vs. 0.111). Thus, the hierarchy not explained solely by institutional size or overall inflow and outflow volume.

We further quantified inequality in the distribution of mobility using Gini coefficients (20) and Herfindahl–Hirschman indices (HHI) (21). Inflows are more concentrated than outflows across prestige tiers (Gini = 0.56 vs. 0.46; HHI = 0.44 vs. 0.35). The same qualitative pattern holds when institutions are partitioned into deciles, and the hierarchy remains stable under a Bradley-Terry paired-comparison model and a size-adjusted inflow-outflow measure (Supplementary Materials S6).

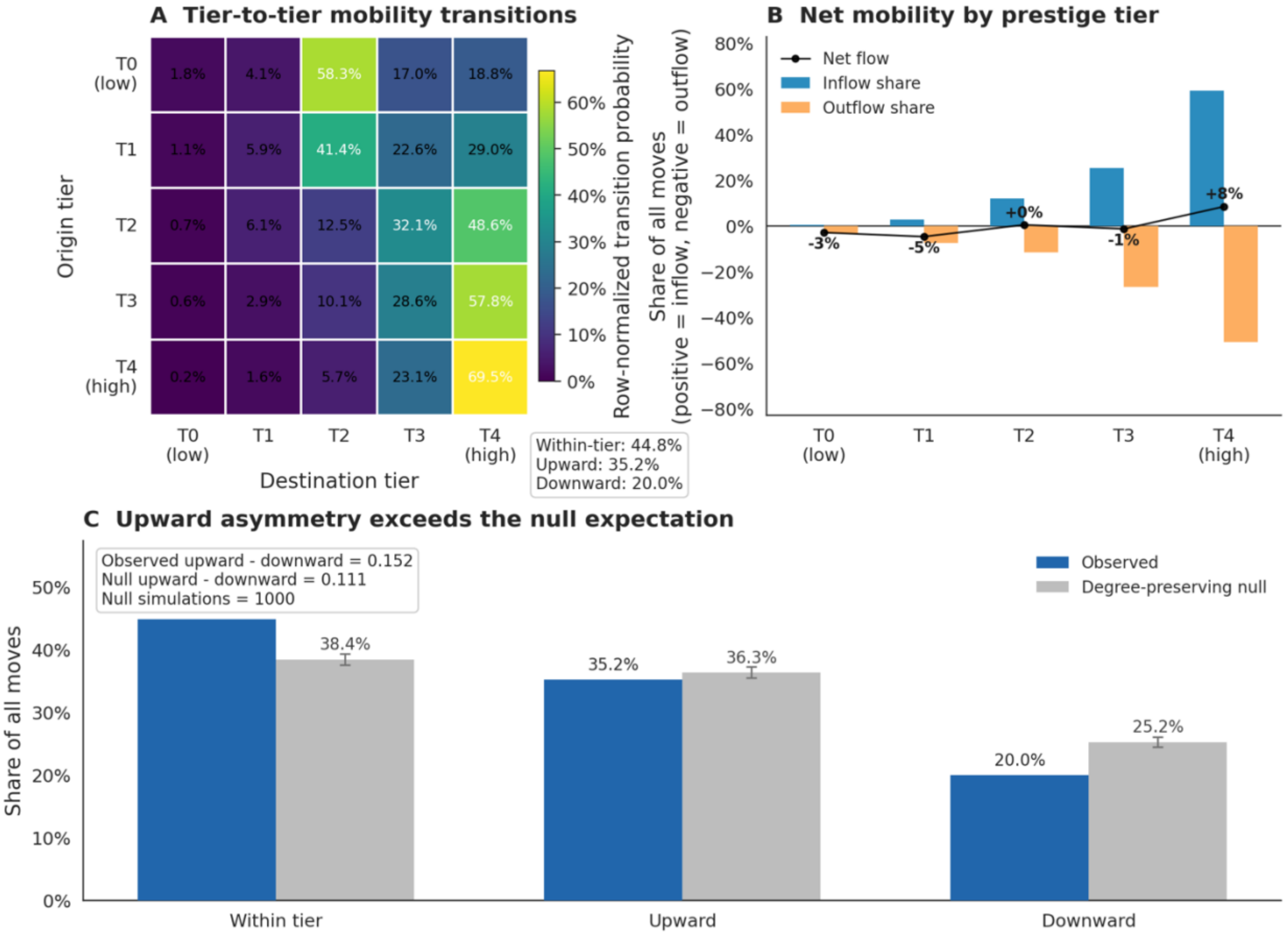

**Fig. 1. Faculty mobility is strongly hierarchical and disproportionately directed toward prestigious institutions.** (A) Row-normalized transition matrix of faculty moves between institutional prestige quintiles derived from PageRank scores computed on the directed mobility network. Rows denote origin tiers and columns denote destination tiers (T0 = lowest prestige; T4 = highest prestige). Diagonal concentration indicates within-tier circulation, while the asymmetric distribution above the diagonal reflects upward mobility toward higher-prestige institutions. (B) Distribution of faculty inflows and outflows across prestige tiers. Positive values indicate inflow shares and negative values indicate outflow shares relative to all observed moves; the black line shows net flow. The highest prestige tier functions as the dominant net importer of faculty, whereas lower tiers are net exporters. (C) Comparison between observed mobility structure and a degree-preserving null model that maintains institutional inflow and outflow volumes. Error bars denote 95% simulation intervals from 1,000 null realizations.

**The mobility hierarchy recapitulates the hiring hierarchy**

We next compare the prestige ordering revealed by career mobility with the ordering observed in faculty hiring and placement. Across institutions observed in both systems, mobility-based and hiring-based PageRank scores are strongly aligned (Spearman rho = 0.75; Fig. 2A). Institutions that dominate initial faculty placement therefore also tend to dominate the later-career mobility market.

The overlap is especially visible near the top of the hierarchy. Among institutions in the top decile of the mobility ranking, 74% are also in the top decile of the hiring ranking, and nearly half of the top 20 institutions appear in both lists. Lorenz curves further show that both systems concentrate inflows among a small share of institutions (Fig. 2B). This alignment suggests that faculty mobility does not create a separate prestige order; rather, it reproduces and extends the institutional hierarchy already established through faculty training, hiring, and placement.

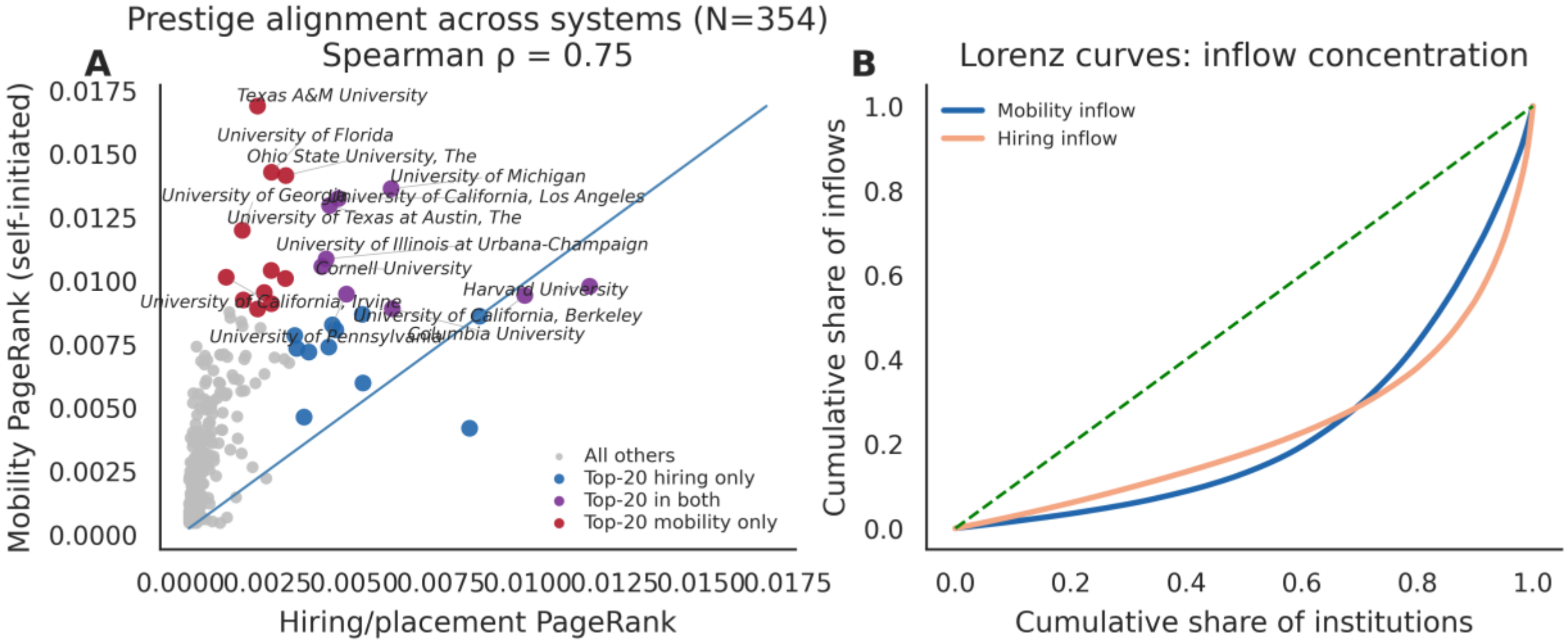

**Fig. 2. Mobility and hiring reveal closely aligned prestige hierarchies.** (A) Relationship between each institution's prestige based on faculty hiring and placement and its prestige based on the academic mobility network. Each point is a U.S. PhD-granting institution observed in both networks (N = 354). The strong rank correlation indicates that institutions that dominate faculty hiring also attract faculty later in their careers. (B) Lorenz curves comparing the concentration of faculty inflows in the hiring/placement network and the mobility network. In both systems, inflows are highly concentrated among a small share of institutions.

**Publication trajectories largely continue along pre-move paths**

We then examine whether relocation is associated with changes in individual research trajectories. We compare observed publication output with trajectories predicted from each mover's pre-move trend. For movers with at least three pre-move observations, we estimate a linear pre-move productivity trajectory and project it forward into the post-move period.

Observed trajectories generally track predictions based on pre-move output (Fig. 3). Upward movers continue to increase output after relocation, but realized output remains below the extrapolated trajectory. Lateral movers show the closest alignment between observed and predicted output. Downward movers display a modest post-move increase followed by stabilization below the predicted trajectory. The same pattern appears across career stages: full professors track predicted trajectories closely, associate professors show small shortfalls, and assistant professors continue to grow but not beyond the growth implied by their pre-move momentum.

These descriptive results do not identify a causal effect of mobility, but they reveal an important feature of the data: moves often occur during ongoing productivity trajectories. Analyses that compare pre- and post-move averages without accounting for those trajectories would therefore risk attributing preexisting momentum to the move itself.

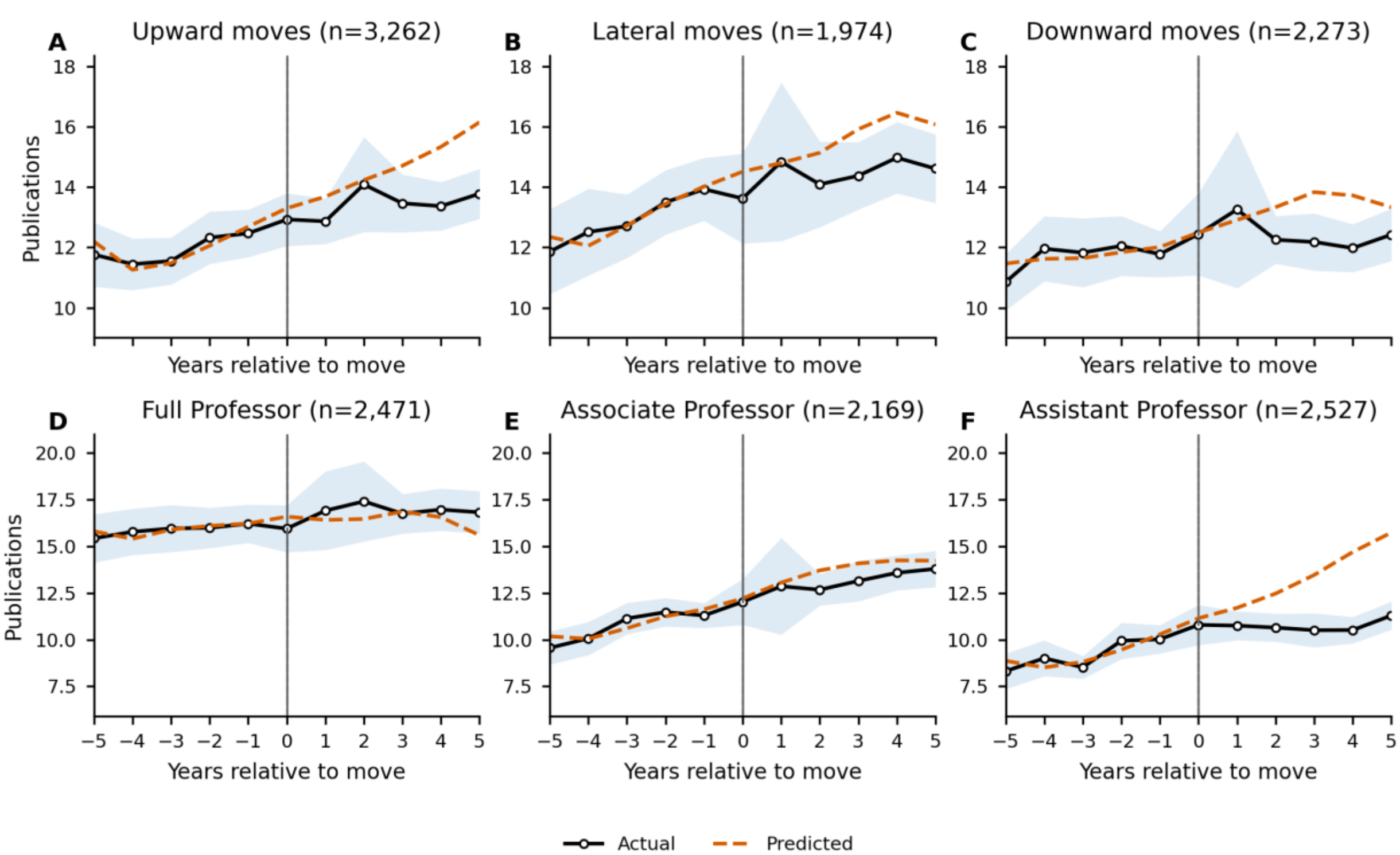

**Fig. 3. Observed publication trajectories track pre-move trends**. Solid lines show observed publication output; dashed lines show output predicted from each faculty member's pre-move productivity trend. Shaded regions indicate 95% confidence intervals. Panels A–C stratify moves by change in institutional prestige; panels D–F stratify movers by academic rank at relocation. The vertical dashed line marks the year of the move.

**Adjusted event studies show limited deviations from expected performance**

To measure deviations from pre-move trajectories more directly, we remove author-specific pre-move linear trends using author-specific linear trends estimated only from pre-move observations and then fit event-study models with author and year fixed effects. This approach asks whether research outcomes depart from the path predicted by a faculty member's own prior trajectory around the time of relocation.

After this adjustment, publication output remains close to predicted levels throughout the post-move period (Fig. 4). Post-move coefficients are small, with average deviations of approximately 0.030 log points and a maximum deviation of 0.048 log points, or about 5%. These estimates provide little evidence of sustained post-move productivity gains, including for moves to more prestigious institutions.

Citation outcomes also show limited and mixed deviations. Field-weighted citation impact declines gradually in later post-move years, reaching about -0.114 log points four years after relocation. The share of publications in the global top 1% of citations follows a similar pattern, declining by about 0.60 percentage points in later post-move years. These changes are modest relative to baseline variation and do not indicate a broad post-move improvement in research impact.

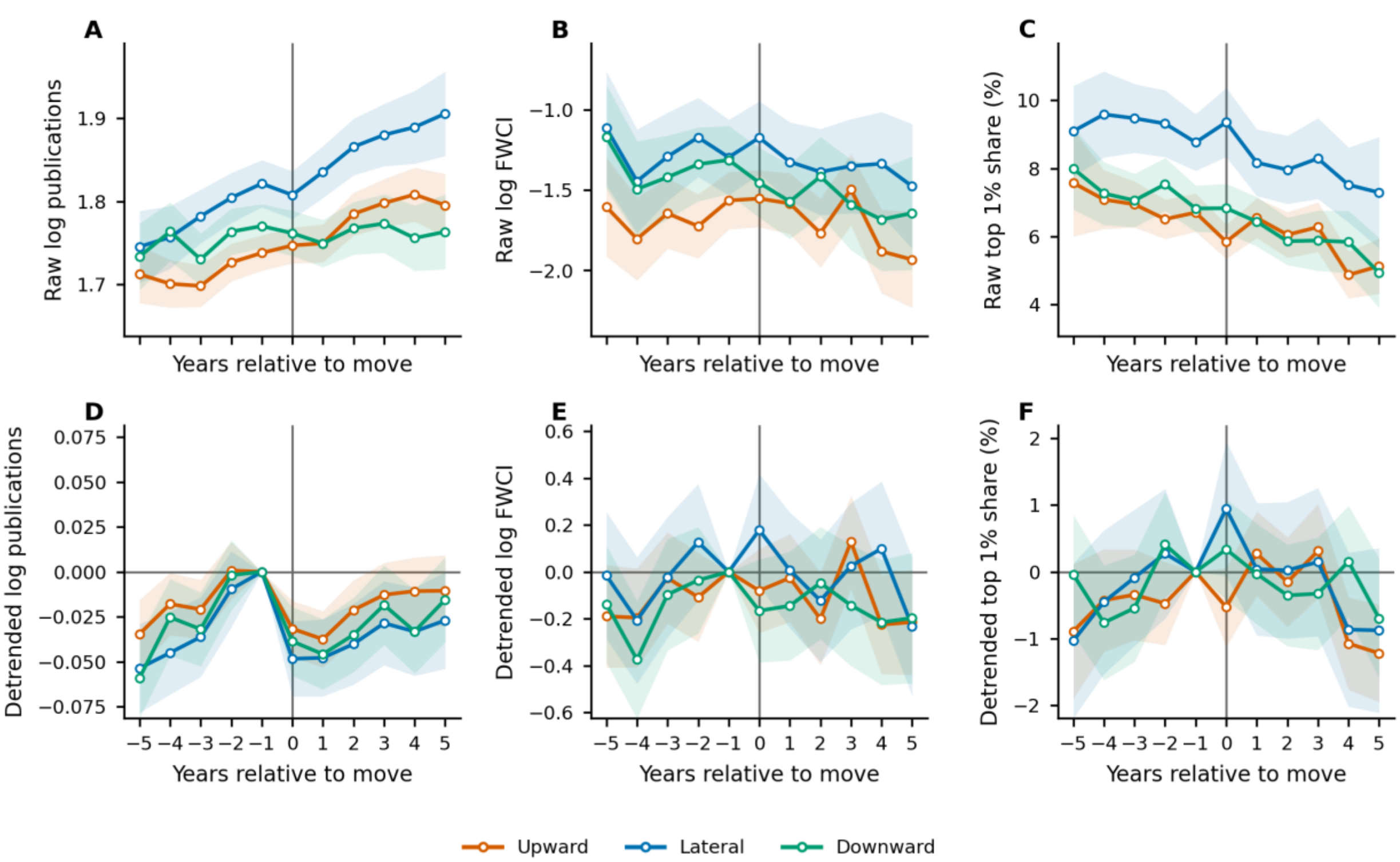

**Fig. 4. Mobility is associated with limited deviations from pre-move research trajectories.** The top row shows raw event-time averages for publication output, citation impact, and top-cited output by move direction. The bottom row shows trajectory-corrected event-study estimates after removing author fixed effects, year fixed effects, author-specific linear pre-trends, and controlling for time-varying academic rank. Coefficients are relative to the year before the move.

**Mobility reorganizes collaboration more than research output**

We next examine whether relocation changes the organization or direction of knowledge production even when output and citation outcomes remain close to prior trajectories. We focus on novelty, disruptiveness, and collaboration.

Research novelty is largely unchanged around the move. Although a small number of event-time coefficients are individually significant, the magnitudes are trivial and a joint test of the event-time coefficients is not statistically significant ($\chi^2 = 8.07$, $p = 0.089$). Average disruptiveness, measured by the CD index, is likewise stable ($\chi^2 = 1.69$, $p = 0.79$). By contrast, the share of papers classified as disruptive rises modestly after relocation ($\chi^2 = 12.17$, $p = 0.016$), suggesting a slightly greater propensity to challenge existing lines of work.

The clearest post-move change occurs in collaboration. The share of new collaborators rises significantly around the mobility event ($\chi^2 = 9.93$, $p = 0.042$). Professors already exhibit slightly elevated levels of new collaboration several years prior to relocation ($\beta = 0.011$ at $t = -4$, $p = 0.016$), consistent with the possibility that expanding collaboration networks precede institutional moves. After relocation, the share of new collaborators increases further, peaking roughly three years after the move ($\beta = 0.018$, $p < 0.001$). These estimates imply that the share of collaborators who are new increases by roughly two percentage points following relocation.

**Robustness and heterogeneity analyses**

Supplementary analyses confirm the robustness of the main findings across multiple specifications and subsamples. Stacked event-study estimates, which compare moving faculty to not-yet-moved and non-moving peers within the same event window, show that research performance evolves smoothly around institutional mobility. Publication output exhibits clear productivity momentum, rising prior to relocation and remaining elevated afterward, with post-move coefficients of approximately +1.2 to +1.8 publications per year relative to the baseline ($F = 4.03$, $p < 0.001$). In contrast, citation impact remains stable relative to non-movers, with citation impact coefficients close to zero and no detectable pre-trends ($\chi^2 = 0.94$, $p = 0.92$). Top-cited output rates fluctuate modestly but show no sustained post-move increase; joint significance ($F = 2.90$, $p = 0.001$) is driven primarily by pre-move variation ($\chi^2 = 15.38$, $p = 0.004$). Overall, these results indicate that, even relative to comparable non-moving professors, research performance largely follows preexisting trajectories rather than shifting in response to relocation.

Stratifying by academic rank reveals no sustained post-move productivity gains for assistant or associate professors, while full professors experience modest productivity declines (≈ 4–5%) accompanied by increases in citation impact (≈ 25–35% five years after the move) (Fig. S2). Discipline-specific analyses show that productivity effects are largely neutral across fields, with modest medium-term increases limited to a few applied domains (Fig. S3; Table S4), and no systematic gains in citation impact or top-cited output rates (Fig. S4–S5). Gender interaction models likewise reveal no statistically significant female–male differences in mobility effects (Fig. S6). Finally, we implement a matched difference-in-differences design that compares movers with observationally similar non-movers from the same department, academic rank, and PhD cohort. Results show that treatment effects are close to zero for productivity ($\beta = -0.006$), citation impact ($\beta = 0.027$), and top-cited output ($\beta = 0.003$), reinforcing the conclusion that institutional mobility does not systematically alter individual research performance.

**Discussion**

Academic mobility is often assumed to raise research output by moving faculty to better-resourced institutions (22-24). Our results suggest a different system-level mechanism. In U.S. research universities, mobility is strongly hierarchical: high-prestige institutions attract disproportionate inflows and lower-prestige institutions experience net losses. Yet movers' publication and citation trajectories remain largely continuous with their pre-move paths. Mobility therefore appears to reallocate research capacity across institutions more than it systematically amplifies individual research output.

Faculty mobility favors already-prestigious institutions. This upward sorting of faculty reinforces cumulative advantage feedback loop: high-prestige institutions accumulate increasing faculty research capacity, while lower-prestige institutions remain at risk of losing productive faculty members. The persistence of a prestige hierarchy suggests that the academic system exhibits a form of advantage lock-in: established institutions with greater reputation, resources, and organizational capacity tend to stay ahead, as they attract talent through both hiring and career moves (4, 30, 31). This pattern aligns with theories of cumulative advantage in science, often known as the "Matthew effect", wherein early successes are indicative of future successes, and initial prestige leads to further prestige (32, 33).

Scientific performance appears highly portable across institutions (10, 25). After accounting for career stage and individual productivity trajectories, professors largely continue along the paths established prior to relocation. Publication output remains close to the predicted trajectory, while citation impact and top-cited output rates exhibit only modest medium-term adjustments. Similarly, the novelty of research and the average disruptiveness of scientific contributions remain stable around mobility events. These patterns are consistent with models in which accumulated human capital, tacit expertise, and established research agendas dominate institutional effects (26). While institutions differ in resources, prestige, and infrastructure, these differences appear to exert limited marginal influence on the productivity trajectories of established professors.

Moves to more prestigious institutions do not yield systematic performance gains. The absence of post-move productivity or citation advantages suggests suggests that institutional prestige may operate more as a sorting signal in faculty labor markets than as a causal driver of subsequent productivity (27). This interpretation helps reconcile a longstanding tension in the literature on academic mobility. Many studies have documented correlations between institutional prestige and scientific output, which are often interpreted as evidence that stronger institutional environments enhance productivity (28). Our results suggest an alternative mechanism: prestigious institutions primarily attract professors who are already highly productive rather than systematically increasing their output after relocation.

Another key finding is that mobility primarily reshapes collaboration networks. Following relocation, faculty members form more new coauthor ties. At the same time, the share of papers classified as disruptive rises modestly, even though the average disruption score remains unchanged. These results suggest that mobility may facilitate new intellectual combinations through the formation of novel collaborative ties. Rather than transforming research agendas or productivity trajectories, institutional moves appear to reconfigure the networks through which ideas are generated and recombined.

Finally, career-stage patterns further clarify the conditions under which productivity portability holds. Assistant professors exhibit steep upward productivity trajectories prior to relocation, but these apparent gains disappear once individual trends are removed, indicating that mobility coincides with rather than causes early-career growth. Associate professors likewise display no systematic post-move deviations in productivity, citation impact, or top-cited output. Among full professors, mobility is associated with modest declines in publication volume but increases in citation impact, suggesting that senior professors may reallocate effort toward fewer but more influential contributions following relocation. These patterns are consistent with adjustment costs associated with rebuilding collaboration networks and organizational capital later in their career (29-31).

More broadly, our findings point to a structural feature of the scientific system: the measured research capacity captured by publication and citation indicators appears to travel largely with faculty members. Mobility therefore functions less as a mechanism for amplifying scientific output than as a process for redistributing scientific talent and reorganizing collaborative networks within the academic ecosystem. Institutional competition for faculty may therefore reallocate measured research capacity across organizations without substantially increasing aggregate publication or citation output.

These findings do not imply that institutions are irrelevant. Institutional contexts shape training, infrastructure, and the formation of early-career networks, particularly during the formative stages of scientific development (32-35). Our results instead suggest that, for faculty in research-intensive universities, the marginal effect of relocating between institutions on research performance is limited. Once professors have developed their research agendas, collaborations, and expertise, these elements appear to be largely portable across institutional environments.

Several limitations should guide interpretation. The analysis is observational; faculty moves are selected events and should not be interpreted as randomized assignments to new institutions. Our trend-adjusted event studies, stacked comparisons, and matched difference-in-differences estimates address several forms of confounding, but unobserved shocks may still coincide with mobility. The analysis focuses on tenure-system faculty at U.S. research universities from 2011 to 2020 and may not generalize to non-tenure-track faculty, non-U.S. systems, or pre-appointment career stages. Finally, bibliometric measures do not capture all dimensions of institutional contribution, including mentoring, teaching, leadership, translational activity, or local capacity building.

## Materials and methods

### Data sources

We use annual faculty rosters from the Academic Analytics Research Center (AARC) spanning 2011–2020 (13, 36). Each annual roster snapshot identifies institutional affiliation, department, academic rank (assistant, associate, full), degree year, and demographic characteristics. Degree year entries equal to 1900 (a placeholder value) or missing were excluded. The sample was restricted to assistant, associate, and full professors. Publications were linked by matching 310,303 U.S.-based AARC faculty to OpenAlex author identifiers (14) using an entity-resolution pipeline. Exact normalized surname blocking reduced the candidate space to 2.46 million comparisons (~8.2 candidates per faculty). Candidate pairs were scored using a transparent composite index combining name agreement, institutional token similarity (average of recall and Jaccard), and binary discipline overlap. Within each faculty member, candidates were ranked and required to exceed calibrated similarity and separation thresholds to ensure dominance over the second-best match. Only high- and medium-confidence matches were retained; these matches account for approximately 81% of all professors in AARC.

### Identification of mobility events

Mobility events were identified by comparing institutional affiliations across consecutive annual snapshots. A move was recorded when the set of employing institutions in year *t* and *t*+1 had no overlap, indicating a complete institutional change. Internal changes, courtesy appointments, and exits from academia were excluded. Only adjacent-year transitions with clear origin and destination were retained. This yields 11,535 mobile professors (12,207 moves) and 208,168 non-mobile professors who were matched to OpenAlex with high or medium confidence and had at least one publication between 2011 and 2020. The unstacked author-year panel aggregates publications and outcomes by faculty member and calendar year. The stacked event-study file expands this panel into person-year-cohort observations by aligning movers and eligible controls to cohort-specific event windows.

**Outcome measures**

We analyze seven author-year outcomes. Productivity is measured as annual publication count and modeled as log(publications + 1). Citation impact is measured as annual mean field-weighted citation impact (FWCI) and modeled as log(FWCI + 0.01). Top-cited output is measured as the annual share of publications in the global top 1% of citations. Novelty is measured as the annual mean novelty score across publications. Disruptiveness is measured using the CD index, and we separately analyze the annual share of papers with positive CD-index values. Collaboration change is measured as the share of coauthors in a given year who have not previously collaborated with the focal author.

**Institutional prestige**

Institutional prestige was measured using PageRank scores computed on the directed network of faculty moves, where nodes denote institutions and edge weights denote observed transitions. For each move $k$, we computed the prestige change as: $\Delta r_k = r_{\{post,k\}} - r_{\{pre,k\}}$ where $r_{\{post,k\}}$ and $r_{\{pre,k\}}$ denote the prestige scores (PageRank values) of the destination and origin institutions, respectively. We then defined a binary upward mobility indicator: $Upward_k = 1[\Delta r_k > 0]$ where $1[\cdot]$ indicates a move to a more prestigious institution, and 0 otherwise. This indicator is based on the continuous PageRank difference between destination and origin institutions, avoiding arbitrary tier thresholds.

Because PageRank is estimated from the same mobility network used to describe directional mobility, we validate the hierarchy in three ways. First, we compare it with independent hiring and placement rankings. Second, we test whether empirical tier transitions exceed a degree-preserving null model that maintains institutional inflow and outflow volumes. Third, we assess temporal stability using PageRank rankings estimated from rolling three-year mobility windows.

**Event-study design**

To examine how institutional mobility is associated with research performance, we employ two complementary event-study designs that balance within-individual identification with external benchmarking. We estimate within-professor event-study models using the sample of 11,535 faculty who change institutions. This design leverages within-person variation to trace changes in multiple outcomes, including productivity, citation impact, top-cited output rates, novelty, disruptiveness, and collaboration, around the mobility event, net of time-invariant individual differences.

For each relocation event, we construct an event-time panel spanning five years before and five years after the move (event time −5 to +5), with the year immediately preceding relocation (event time −1) serving as the reference period. The baseline specification is:

$$y_{it} = \sum_{k \neq -1} \beta_k D_{it}^k + \alpha_i + \gamma_t + \epsilon_{it}$$

where $y_{it}$ denotes the outcome for professor $i$ in year $t$. Outcomes include annual publication output, field-weighted citation impact (FWCI), the count of top 1% publications, novelty, CD index, and share of new collaborators. $D_{it}^k$ is an indicator for event time $k$, $\alpha_i$ are author fixed effects that absorb time-invariant differences across professors, and $\gamma_t$ are year fixed effects capturing common system-wide shocks. Standard errors are clustered at the author level.

To address potential selection on trends, we removed author-specific linear pre-trends. For each author: $Y_{it} = a_i + b_i t + u_{it}$ and outcomes were residualized: $\widetilde{Y_{it}} = Y_{it} - (a_i + b_i t)$. Event-study regressions were then estimated on detrended outcomes (37). Academic rank was included as a time-varying control. Standard errors were clustered at the author level. Domain-specific models were estimated by discipline using the same fixed-effects framework. Gender heterogeneity was assessed via interaction models including event-time × Female indicators.

To assess career-stage heterogeneity, we stratified movers by pre-move academic rank (assistant, associate, full professor) and estimated rank-specific event studies. To isolate deviations from underlying career dynamics, we first removed individual linear productivity trends estimated over the pre-move period. The detrended outcomes were then transformed using a two-way fixed-effects procedure that subtracts author and year means. Event-time coefficients were computed as the mean of the residualized outcome at each relative year, with standard errors derived from the cross-sectional dispersion of residuals within event-time bins.

**Stacked event-study comparison**

As an external benchmark, we implement a stacked event-study design that combines movers with a comparison group of 208,168 non-moving faculty, allowing us to benchmark movers against not-yet-moved peers within the same event window. Observations are restricted to event times -5 through +5. This design reduces bias from staggered treatment timing and provides a counterfactual trajectory for core outcomes: publication output, citation impact, and top-cited output.

**Matched difference-in-differences**

As a robustness check, we implemented a matched DiD design (38). For each mover with move year $t_0$, we define a pre-move window spanning $t_{0-3}$ to $t_{0-1}$. Movers are matched to three non-movers using exact department, exact academic rank, PhD cohort within ±2 years, and minimum distance in pre-move log productivity. Controls are not reused. This procedure yields 6,837 matched movers and 20,511 unique controls.

For each individual, we compute the change in the outcome from the pre-period ($t_{0-3}$ to $t_{0-1}$) to the post-period ($t_{0+1}$ to $t_{0+3}$), excluding the move year. The difference-in-differences estimate is obtained by regressing this individual-level change on an indicator for mover status with HC1 heteroskedasticity-robust standard errors.

**Conflict of interest statement**
The authors declare no financial conflict of interest.

**Author contributions**
EY: Conceptualization, Methodology, Software. EY: Data curation. EY and CN: Writing - Original draft preparation. EY and CN: Visualization, Investigation.

**AI use statement**
Portions of code development and text editing were assisted using a large language model (ChatGPT 5.5). All outputs were reviewed, validated, and edited by the authors.

**Data availability**
AARC roster data are subject to data-use restrictions and cannot be redistributed by the authors. OpenAlex data are publicly available. Code and the aggregate data necessary to reproduce the figures and tables are available at GitHub at https://github.com/erjiayan/faculty-mobility and at Zenodo at https://doi.org/10.5281/zenodo.19712199.

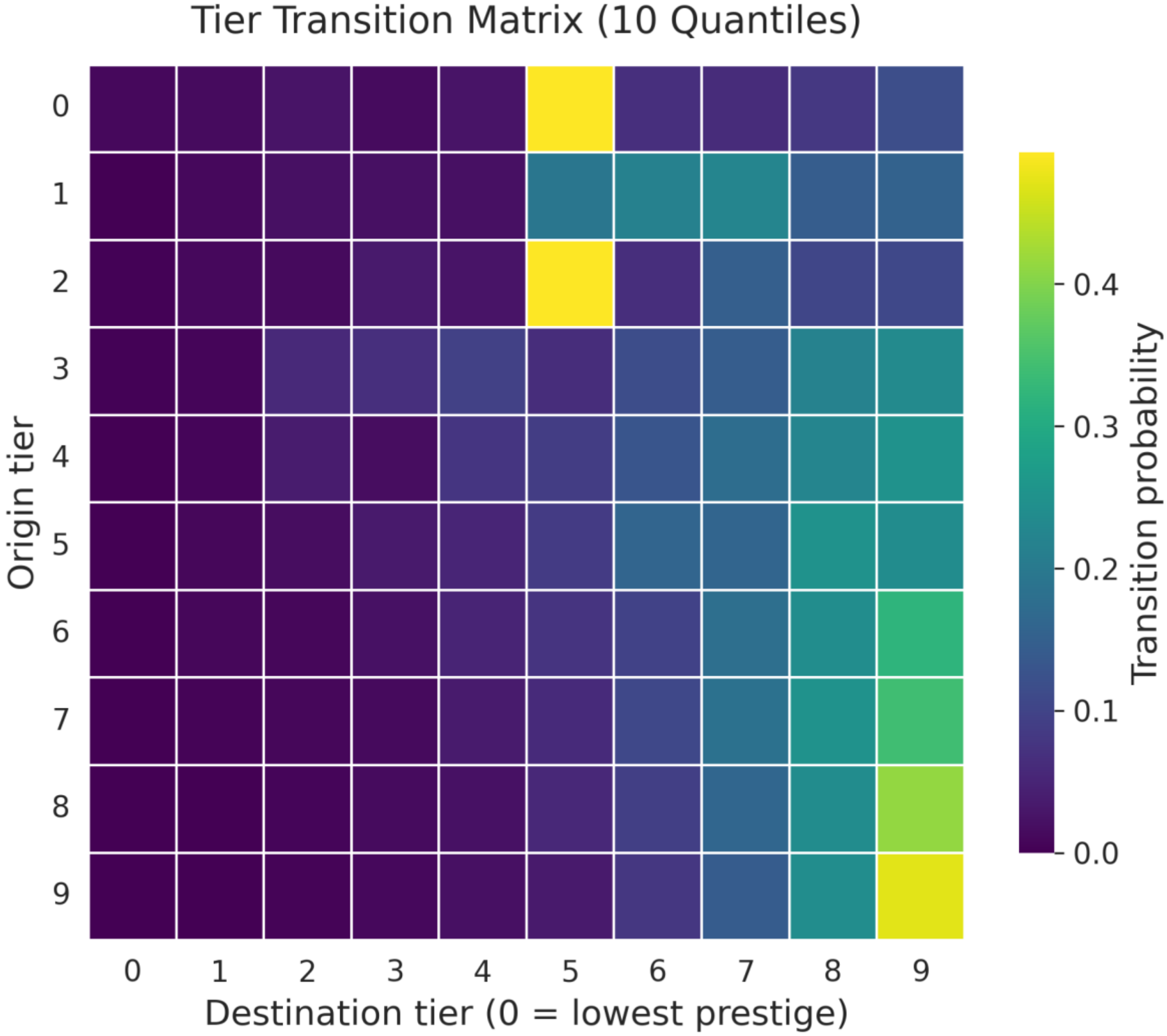

**Fig. S1. Faculty mobility across institutional prestige deciles.** Using prestige deciles rather than quintiles produces more granular transition patterns while preserving the same hierarchical pattern: dominant within-tier exchange and systematic net upward mobility toward the highest deciles. Results confirm that the observed stratification is not an artifact of tier aggregation.

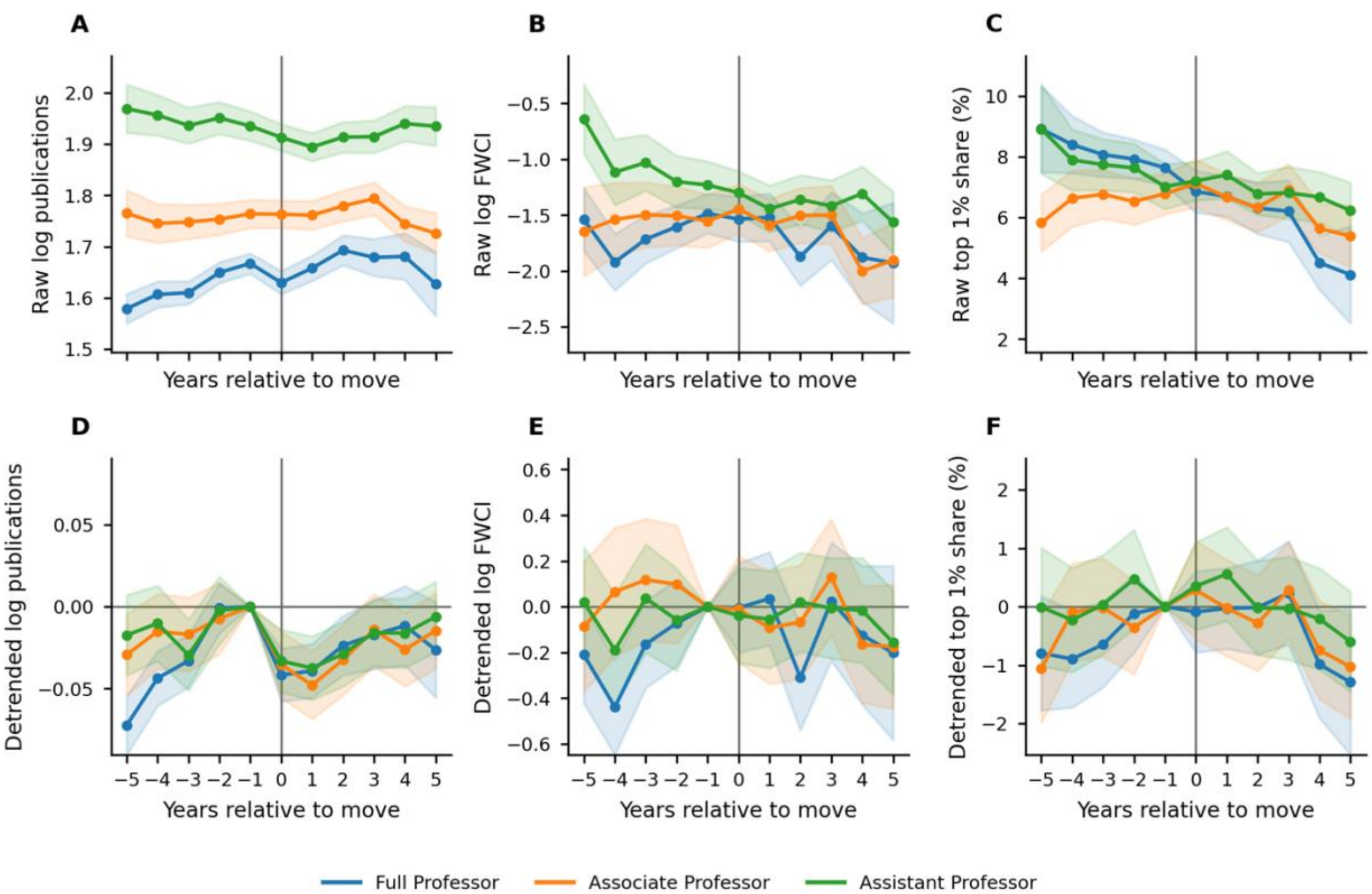

**Fig. S2. Career-stage heterogeneity in research trajectories around institutional mobility.** The top row shows raw event-time averages of research productivity (log publications), citation impact (log field-weighted citation impact, FWCI), and top-cited output (top 1% publication rate), disaggregated by pre-move academic rank (assistant, associate, full professor). Shaded areas represent 95% confidence intervals. Vertical line denotes the move year (event time 0). Bottom row presents fully controlled event-study estimates after removing author fixed effects, year fixed effects, and author-specific linear productivity trends estimated over the pre-move period and controlling for time-varying academic rank. Coefficients are relative to event time −1. Across career stages, mobility does not generate sustained gains in publication output.

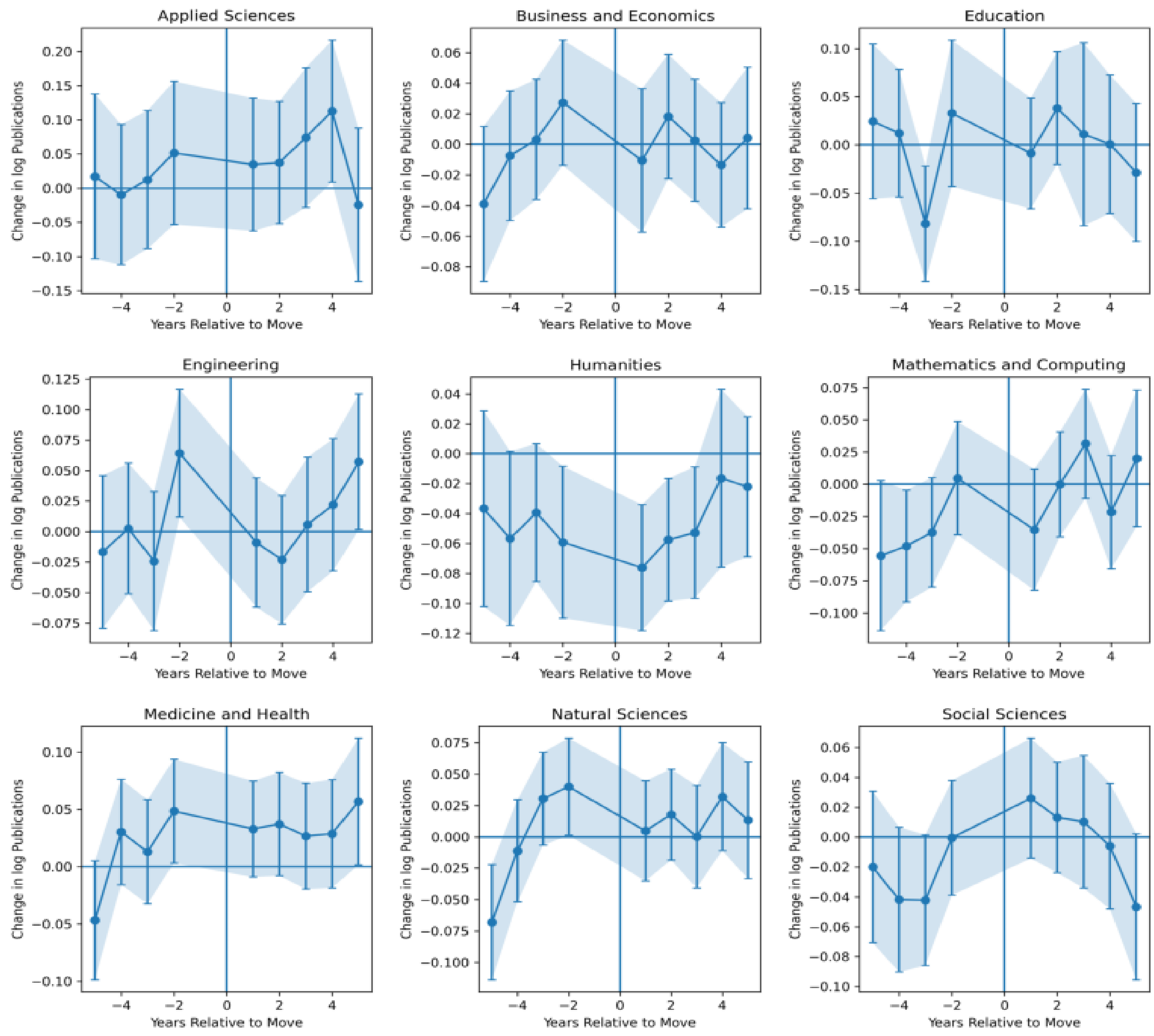

**Fig. S3. Discipline-specific productivity dynamics following faculty mobility.** Event-study estimates of log publication output by umbrella discipline relative to the year prior to mobility (event time −1 omitted). All models include author fixed effects, year fixed effects, and time-varying academic rank controls; standard errors are clustered at the author level. Shaded regions indicate 95% confidence intervals. After full controls, discipline-level heterogeneity attenuates substantially. Only Applied Sciences ($\beta_{+4} \approx 0.113$, $p < 0.05$), Engineering ($\beta_{+5} \approx 0.057$, $p < 0.01$), and Medicine & Health ($\beta_{+5} \approx 0.057$, $p < 0.01$) display modest medium-term productivity gains.

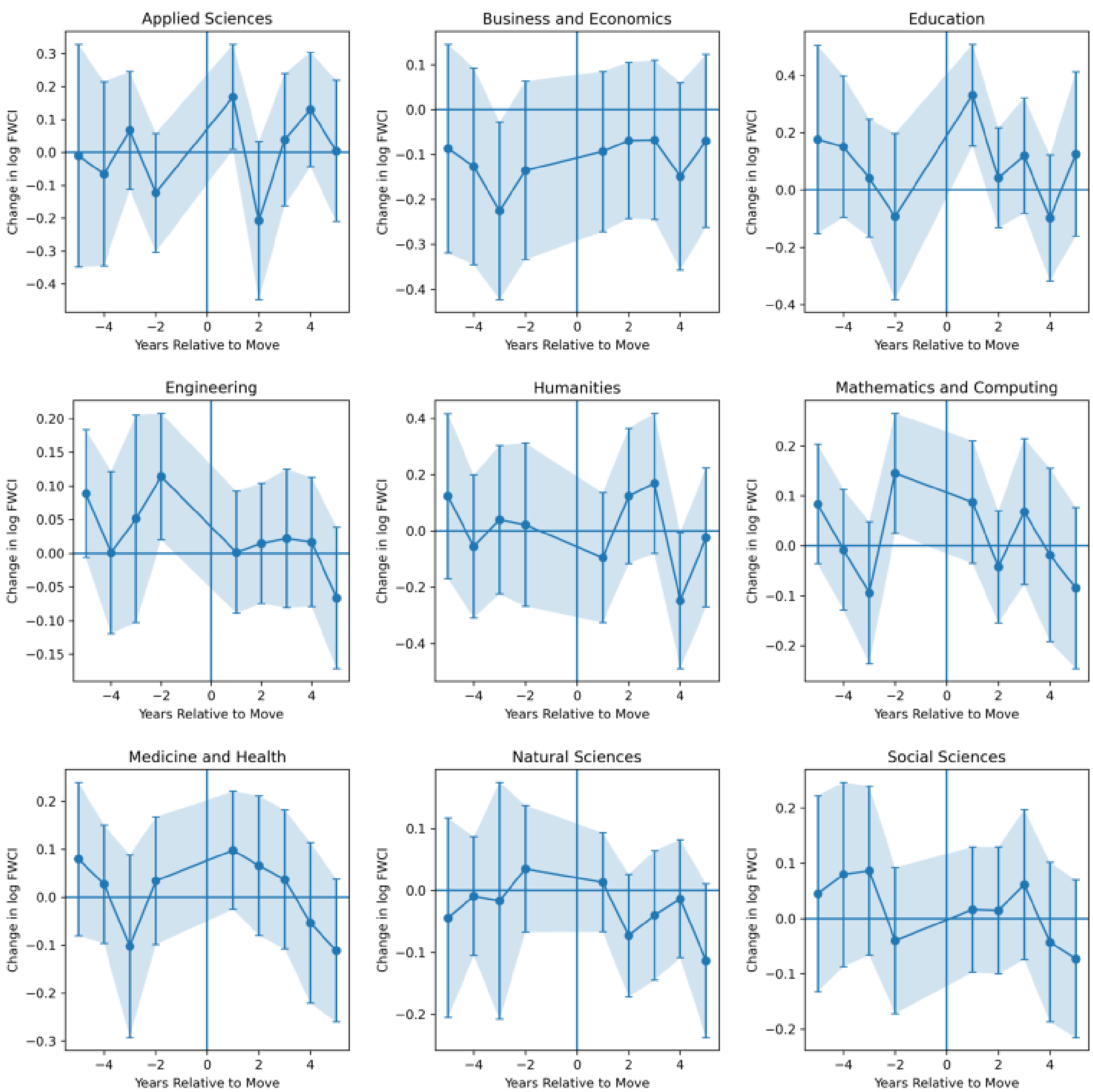

**Fig. S4. Domain-level citation impact dynamics.** Event-study estimates of log FWCI by umbrella discipline. Models include author FE, year FE, and time-varying rank controls. Across all disciplines, post-move coefficients are small and statistically insignificant, indicating no systematic citation impact gains following mobility. Laboratory-intensive and applied disciplines alike show fluctuating but non-persistent effects.

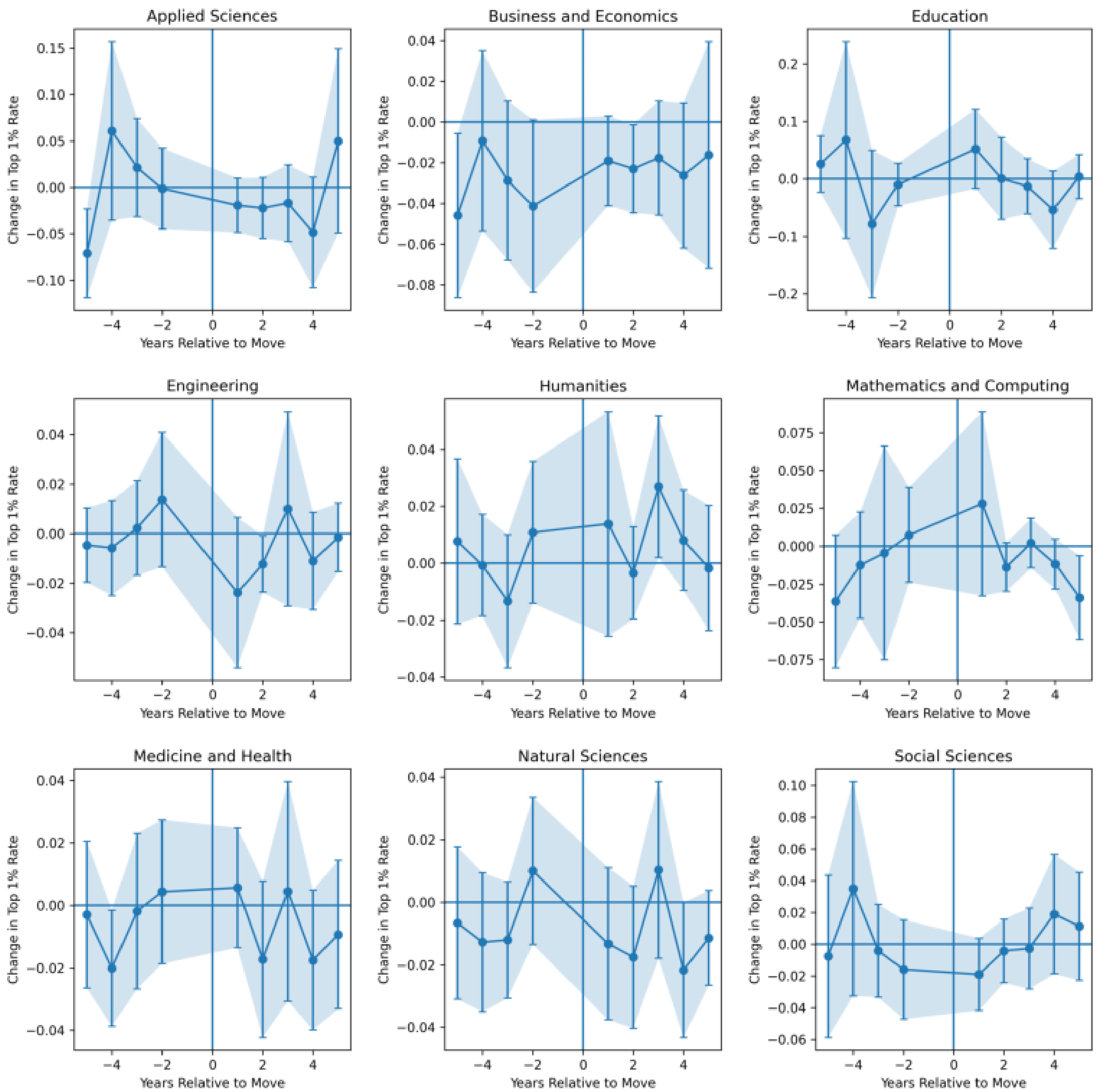

**Fig. S5. Domain-level top-cited output dynamics (Top 1% rate).** Event-study estimates of top 1% publication probability by umbrella discipline under full controls. Coefficients are uniformly small and statistically indistinguishable from zero across disciplines. No discipline exhibits a sustained increase in breakthrough publication probability following mobility.

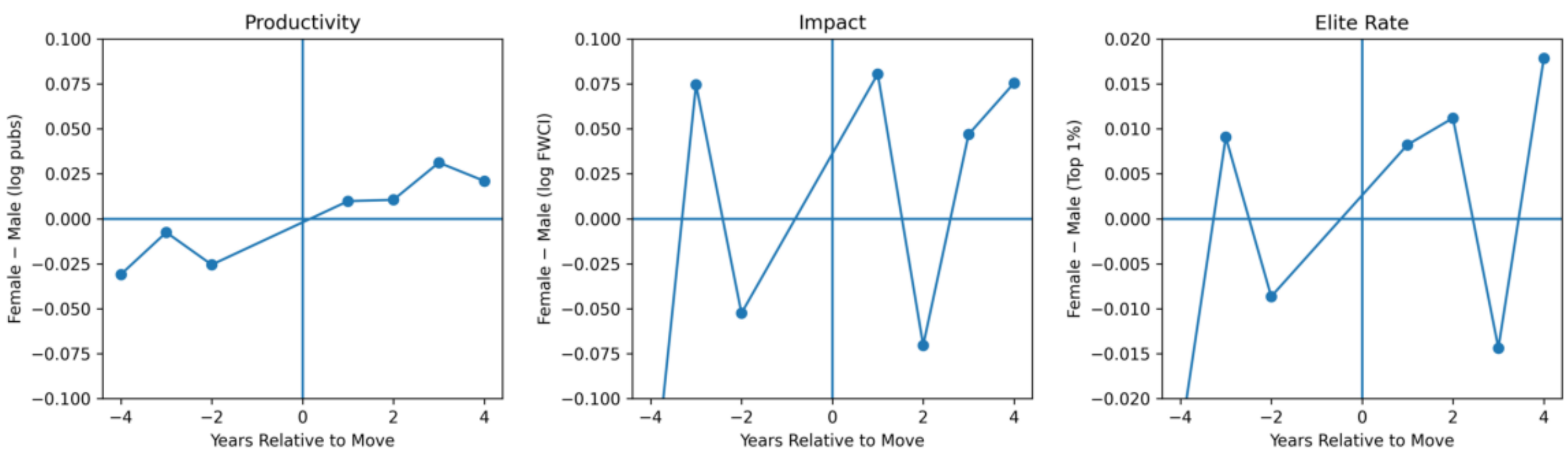

**Fig. S6. Gender-differential mobility effects across productivity, impact, and top-cited output.** Event-study estimates of the difference in mobility effects between women and men on log publication output, log citation impact (FWCI), and top 1% publication rate. Models include author fixed effects, year fixed effects, and time-varying rank controls; standard errors are clustered at the author level. Baseline period is event_time −1. Across all three outcomes, female–male differential effects remain small, indicating no systematic gender-specific mobility effects.

**Supplementary Table S1. Academic mobility flows and concentration metrics by prestige tier** (Inflow Gini = 0.56, Outflow Gini = 0.45, Inflow HHI = 0.44, Outflow HHI = 0.35). For each prestige quintile (Tier 0 = lowest prestige to Tier 4 = highest prestige), the total number of faculty departures (outflows) and arrivals (inflows) is listed, along with the net flow (inflows minus outflows) and that tier's share of all moves. The top prestige tier captures 60% of all incoming faculty moves and is the only group with a large positive net gain, whereas the lowest tiers are net exporters of faculty.

| Tier | Outflow | Inflow | Net flow | Inflow share | Outflow share |
|---|---|---|---|---|---|
| 0 | 393 | 57 | -336 | 0.005 | 0.034 |
| 1 | 870 | 329 | -541 | 0.029 | 0.075 |
| 2 | 1344 | 1398 | 54 | 0.121 | 0.117 |
| 3 | 3074 | 2923 | -151 | 0.254 | 0.267 |
| 4 | 5844 | 6818 | 974 | 0.592 | 0.507 |

**Supplementary Table S2. Tier-specific mobility flows and concentration under decile-based prestige hierarchy (Inflow Gini = 0.58, Outflow Gini = 0.47, Inflow HHI = 0.22, Outflow HHI = 0.18).** Institutions were ranked using weighted PageRank derived from the directed network of faculty moves and partitioned into ten equal-frequency prestige tiers (deciles). For each tier, we report total outflows, inflows, net flow (inflow − outflow), and shares of moves. Consistent with the quintile-based analysis in the main text, flows exhibit strong hierarchical asymmetry: the highest decile functions as a net importer of faculty and captures a disproportionate share of inflows, whereas the lowest deciles act primarily as net exporters.

| Tier | Outflow | Inflow | Net flow | Inflow share | Outflow share |
|---|---|---|---|---|---|
| 0 | 286 | 12 | -274 | 0.001 | 0.025 |
| 1 | 107 | 45 | -62 | 0.004 | 0.009 |
| 2 | 499 | 109 | -390 | 0.009 | 0.043 |
| 3 | 371 | 220 | -151 | 0.019 | 0.032 |
| 4 | 474 | 320 | -154 | 0.028 | 0.041 |
| 5 | 962 | 1117 | 155 | 0.097 | 0.083 |
| 6 | 1315 | 1162 | -153 | 0.101 | 0.114 |
| 7 | 1727 | 1782 | 55 | 0.155 | 0.15 |
| 8 | 2270 | 2653 | 383 | 0.23 | 0.197 |
| 9 | 3514 | 4105 | 591 | 0.356 | 0.305 |

**Supplementary Table S3. Stability of institutional prestige rankings over time.** Spearman rank correlations of university prestige scores computed from 3-year rolling mobility windows. Each entry shows the rank correlation between two time windows (e.g., 2012–2014 vs. 2014–2016). Prestige ordering is highly stable: correlations between adjacent windows consistently exceed $\rho = 0.90$ and remain above $\rho \approx 0.80$ even across multi-year gaps. These results indicate that the prestige hierarchy reflects enduring structural characteristics rather than short-term fluctuations, consistent with a cumulative-advantage process.

| | 2012-2014 | 2013-2015 | 2014-2016 | 2015-2017 | 2016-2018 | 2017-2019 | 2018-2020 |
|---|---|---|---|---|---|---|---|
| 2012-2014 | 1.00 | 0.93 | 0.88 | 0.83 | 0.84 | 0.83 | 0.78 |
| 2013-2015 | 0.93 | 1.00 | 0.92 | 0.89 | 0.85 | 0.84 | 0.80 |
| 2014-2016 | 0.88 | 0.92 | 1.00 | 0.96 | 0.91 | 0.86 | 0.82 |
| 2015-2017 | 0.83 | 0.89 | 0.96 | 1.00 | 0.94 | 0.89 | 0.82 |
| 2016-2018 | 0.84 | 0.85 | 0.91 | 0.94 | 1.00 | 0.96 | 0.90 |
| 2017-2019 | 0.83 | 0.84 | 0.86 | 0.89 | 0.96 | 1.00 | 0.94 |
| 2018-2020 | 0.78 | 0.80 | 0.82 | 0.82 | 0.90 | 0.94 | 1.00 |

**Supplementary Table S4. Fully controlled domain-specific productivity effects.**

| **Domain** | **Most Negative Pre (β)** | **Event year** | **Largest Post (β)** | **Event year** |
|---|---|---|---|---|
| Applied Sciences | −0.009 | −4 | 0.113* | 4 |
| Business & Econ | −0.039 | −5 | 0.018 | 2 |
| Education | −0.082* | −3 | 0.038 | 2 |
| Engineering | −0.024 | −3 | 0.057** | 5 |
| Humanities | −0.059* | −2 | −0.016 | 4 |
| Mathematics & Computing | −0.055* | −5 | 0.031 | 3 |
| Medicine & Health | −0.047 | −5 | 0.057** | 5 |
| Natural Sciences | −0.068* | −5 | 0.032 | 4 |
| Social Sciences | −0.042 | −3 | 0.026 | 1 |

Notes: All models include author FE, year FE, and rank controls. Baseline = −1. Standard errors clustered at author level. *: * $p < 0.05$; ** $p < 0.01$

# Supplementary materials

## S1. Data construction

### Faculty census and institutional structure

We use annual faculty rosters from the Academic Analytics Research Center (AARC) for the years 2011–2020 (13, 36). Each annual capture (AADVersion) identifies the institution, department (UnitId), academic rank (RankTypeId), degree year, and demographic characteristics of faculty members. DegreeYear entries equal to 1900 or missing were excluded to avoid placeholder values. RankTypeId was restricted to assistant, associate, and full professors.

### Publication linkage

We developed a scalable, reproducible entity-resolution pipeline to link 310,303 U.S.–based AARC faculty to OpenAlex author IDs (14), framing the task as a one-to-one matching problem between ~310k faculty and 100M+ authors while minimizing false positives and false negatives. To avoid infeasible full pairwise comparisons, we applied restrictive surname-based blocking, reducing the search space to 2,460,398 candidate pairs (~8.16 candidates per faculty), after testing and rejecting looser blocking schemes that generated severe skew for high-frequency surnames. For each candidate pair, we computed a transparent composite score ($S = S_{name} + 0.3S_{inst} + 0.25S_{affil} + 0.1S_{discipline} + 0.1S_{prod}$), comprising name blocking strength, affiliation overlap, institutional similarity, discipline consistency, and productivity alignment, with greater emphasis on affiliation and institutional signals.

Candidates were ranked per faculty member, and we defined a separation metric ($Delta = S_1 - S_2$) to quantify ambiguity, requiring dominance of the top match over the runner-up to mitigate common-surname inflation. Thresholds were empirically calibrated and shown to be robust to small perturbations (±0.02 changes altered the high-confidence tier by <3%): High confidence was assigned when ($S \geq 0.80$) and ($Delta \geq 0.05$); medium confidence was assigned when ($0.50 \leq S < 0.80$) and ($Delta \geq 0.01$); otherwise matches were classified as Low. Approximately 2.8% of faculty had no viable OpenAlex candidate. The overall design favors precision over aggressive recall, emphasizing interpretability, dominance-based disambiguation, and robustness to mobility, interdisciplinary careers, and high-frequency surnames.

## S2. Metrics

Faculty were linked to OpenAlex author identifiers using a validated crosswalk with confidence-tier filtering (High or Medium only). The resulting author–paper dataset contains over 200,000 professors, including publication year, FWCI, top 1% and top 10% citation indicators, authorship position, novelty score, CD index, and share of new collaborators. Author–year panels were constructed by aggregating publications per author per year:

- **Productivity**: total publications per year; log(pubs + 1) used in regression.
- **Citation impact**: mean FWCI per year; log(FWCI + 0.01) transformation.
- **Top-cited output**: top 1% publication rate = number of top 1% publications / total publications.
- **Novelty** captures the extent to which a publication combines previously unconnected areas of knowledge. For each paper, we examine pairwise combinations of subject categories among its cited references and measure how frequently those combinations appeared in the prior literature. Rare combinations receive higher novelty scores, indicating new intellectual linkages across domains. Author–year novelty is calculated as the mean novelty score across all publications produced by a professor in a given year.

- **Disruptiveness** is measured using the simplified CD index, which evaluates whether a publication redirects subsequent research away from its cited predecessors. For a focal paper i, later citing papers are classified into those that cite the focal paper without citing its references ($N_i$) and those that cite both the focal paper and at least one of its references ($N_j$). The CD index is defined as: $CD_i = (N_i - N_j)/(N_i + N_j)$. We use the simplified CD index, which omits the Nk term. This standard approximation in large-scale citation analyses yields nearly identical disruption rankings while greatly reducing computational burden. In our setting, more than 13 million works are associated with the 220,000 professors observed during the study period, producing more than 1.8 billion citation links (39, 40). Positive values indicate disruptive research that displaces prior work, whereas negative values indicate developmental research that consolidates existing knowledge. Author–year disruptiveness is measured as the mean CD index across all publications produced in that year.
- **Share of disruptive publications**: To reduce noise in the CD index, we also construct a binary measure indicating whether a paper is disruptive ($CD_i > 0$). The share of disruptive publications is the fraction of a professor's publications in a given year with positive disruption scores.
- **New collaborator share**: To measure changes in collaboration networks, we compute the share of coauthors in a given year who have not previously collaborated with the focal professor. The new collaborator share equals the fraction of collaborators in that year who are newly observed coauthors, capturing the extent to which professors expand their collaboration networks over time.

## S3. Mobility identification

We identify faculty mobility events by comparing each faculty member's institutional affiliation across consecutive annual snapshots. For each individual, we construct the set of employing institutions in year *t* and in year *t*+1; a mobility event is flagged when these two sets have no overlap, indicating a complete change of institutional affiliation. We exclude internal changes within the same university, courtesy or joint appointments, and exits from academia. Exits are treated as right-censored observations rather than as moves. To ensure continuity, we restrict attention to transitions with a clear origin and destination in adjacent years. Rare cases involving gap years or temporary absences are handled conservatively to avoid misclassifying sabbaticals or temporary leaves as permanent moves. Each retained mobility event is recorded as a directed origin–destination pair of institutions. This yields 11,535 mobile professors (12,207 moves) and 208,168 non-mobile professors who can be matched in OpenAlex with high or medium confidence matches and having at least one publication between 2011 and 2020.

## S4. Event-study identification

### Baseline specification

We estimate:

$$y_{it} = \sum_{k \neq -1} \beta_k D_{it}^k + \alpha_i + \gamma_t + \epsilon_{it}$$

where $y_{it}$ denotes the outcome for professor *i* in year *t*. Outcomes include annual publication output, field-weighted citation impact (FWCI), the count of top 1% publications, novelty, CD index, and share of new collaborators. $D_{it}^{k}$ is an indicator for event time $k$, $\alpha_i$ are author fixed effects that absorb time-invariant differences across professors, and $\gamma_t$ are year fixed effects capturing common system-wide shocks. Standard errors are clustered at the author level. This removes all time-invariant individual heterogeneity (37). In the stacked specification, movers are compared to professors who have not yet moved at the same relative event time, reducing bias from staggered treatment timing.

**Adjustment for pre-trend bias**

Because mobility may be preceded by performance decline or growth, we remove author-specific linear trends prior to estimation. For each author *i*, we estimate:

$$Y_{it} = a_i + b_i t + u_{it}$$

Where:

- $a_i$ is the individual-specific intercept
- $b_i$ is the individual-specific linear time trend

and compute residualized outcomes:

$$\widetilde{Y_{it}} = Y_{it} - (a_i + b_i t)$$

Where:

- $\widetilde{Y_{it}}$ is the detrended outcome
- $Y_{it}$ is the observed outcome

Event-study regressions are then performed on $\widetilde{Y_{it}}$, isolating deviations from predicted trajectory. This procedure reduces bias arising from linear career-stage selection into mobility.

**Time-varying controls**

Academic rank (RankTypeId) is included as a time-varying control to account for concurrent promotions. Because discipline is largely time-invariant at the individual level and citation impact (FWCI) is field-normalized, discipline effects are absorbed by author fixed effects.

**Prestige measure**

Institutional prestige is measured using PageRank-based quantile blocks (19) derived from the directed faculty mobility network. Prestige change is defined as:

$$\Delta\text{prestige} = \text{Block}_{\text{destination}} - \text{Block}_{\text{origin}}$$

Moves are classified as upward, lateral, or downward. Event-study regressions are estimated separately by move type to test whether performance effects depend on institutional prestige change.

## S5. Difference-in-differences

Institutional mobility is potentially selective: faculty who move may differ systematically from non-movers in observable and unobservable characteristics. Although the main text employs an event-study design with author fixed effects and individual linear trends to address selection and differential trajectories, we implement a matched difference-in-differences (DiD) robustness check to further assess whether post-move performance patterns are attributable to mobility rather than preexisting differences (38). This matched DiD approach directly compares movers to observationally similar non-movers within the same departmental and career contexts.

**Definition of treatment and control groups**
The pool of faculty is defined using the full author-year panel, linked to faculty identifiers. Mobility status is determined by comparing each faculty member's institutional affiliation across consecutive annual snapshots. The full sample includes 11,535 movers. Of these, 6,837 are successfully matched under strict criteria described below.

**Matching procedure**
For each mover *i* with move year $t_0$, we define a pre-move window spanning $t_{0-3}$ to $t_{0-1}$. Baseline characteristics are computed over this window, including:

- Average log publication output: log(pubs + 1)
- Average log field-weighted citation impact (FWCI): log(FWCI + 0.01)
- Average top-cited output rate (share of top 1% publications).

Movers are matched to three nearest-neighbor non-movers satisfying:

- Exact match on department (UnitId).
- Exact match on academic rank (RankTypeId).
- PhD cohort proximity within ±2 years.
- Minimum distance in pre-move log productivity.

To reduce dependence and over-representation, controls are used at most once (no reuse). This greedy nearest-neighbor algorithm yields 20,511 unique controls (three per matched mover), producing a matched sample of 27,348 faculty. Unmatched movers typically belong to small departments or rare cohorts for which no comparable non-movers exist under strict matching constraints.

**Construction of pre/post windows**
For each matched faculty member, we define:

- Pre period: $t_{0-3}$ to $t_{0-1}$.
- Post period: $t_{0+1}$ to $t_{0+3}$.

The move year $t_0$ is excluded to avoid transitional effects.

For each faculty member *i*, we compute:

$$\Delta Y_i = \bar{Y}_{post,i} - \bar{Y}_{pre,i}$$

Where:

- $\bar{Y}_{post,i}$ is the average outcome for individual *i* in the post-period
- $\bar{Y}_{pre,i}$ is the average outcome for individual *i* in the pre-period
- log(pubs + 1)
- log(FWCI + 0.01)
- top 1% publication rate

This differencing removes individual fixed heterogeneity and yields an author-level change measure.

**Difference-in-differences estimator**
The matched DiD estimator is obtained from:

$$\Delta Y_i = \beta \cdot Mover_i + \epsilon_i$$

- $\Delta Y_i$ is the pre–post change in the outcome for individual $i$
- $Mover_i$ is the indicator equal to 1 if individual *i* moved, 0 otherwise

This specification is equivalent to a two-period fixed-effects DiD model:
$Y_i = \alpha_i + \gamma_t + \beta(Mover_i \times Post_t) + \epsilon_i$ with individual fixed effects differenced out. Standard errors are estimated using heteroskedasticity-robust (HC1) variance.

**DiD results**
Across 27,348 matched faculty, post-move performance differences are small and statistically indistinguishable from zero:

- Productivity: $\beta$ = -0.006 (SE = 0.008, t = −0.85)
- Citation impact: $\beta$ = +0.027 (SE = 0.025, t = 1.10)
- Top-cited output: $\beta$ = +0.003 (SE = 0.003, t = 0.85)

These magnitudes correspond to:

- <1% change in publication output,
- ~2–3% change in citation impact,
- <0.3 percentage-point change in top-cited output.

None of the estimates is statistically significant at conventional levels.

The matched DiD results confirm that institutional mobility is not associated with systematic gains or losses in research productivity, citation impact, or top-cited output once observable selection and baseline differences are accounted for. In contrast to the event-study specification, which reveals modest medium-term deviations relative to individual performance trajectories, the matched DiD suggests that such deviations are not large relative to comparable non-movers in the same institutional and cohort contexts.

The matched DiD design imposes strict matching constraints and excludes approximately 46% of movers without comparable controls under these criteria. While this enhances internal validity, it may reduce external generalizability to movers in smaller departments or rare cohorts. Nevertheless, the matched DiD results are consistent with the full-sample event-study findings.

## S6. Mobility hierarchy robustness

We verified that our mobility hierarchy findings are robust to alternative implementations. First, we replicated the tiered mobility analysis using deciles (10 equal-sized tiers) instead of quintiles. The key patterns held: 45.4% of moves were upward, the top decile captured 36% of all inflows with a net gain of +591 faculty, and inequality metrics were even more extreme (inflow Gini = 0.58; HHI = 0.22), indicating intensified concentration at the top. Second, we tested alternative prestige metrics, including a Bradley–Terry paired-comparison model and a size-adjusted inflow–outflow ratio. Both alternatives yielded tier structures closely aligned with PageRank (Spearman ρ = 0.48 and 0.63, respectively), supporting the validity of the prestige ordering. Third, we benchmarked against degree-preserving null models that maintain each institution's inflow and outflow volumes. The empirical mobility matrix exhibits stronger within-tier sorting (44.8% vs. 38.4%) and a more pronounced upward asymmetry (35.2% upward vs. 20% downward; Δ = 0.152) than expected under the null. Although the null model itself displays mild upward bias (36.3% vs. 25.3%; Δ = 0.111), the empirical system amplifies this asymmetry substantially, indicating that prestige-stratified mobility cannot be explained by degree constraints alone and instead reflects additional hierarchical structure. Finally, we assessed the temporal stability of institutional prestige by computing rank correlations across PageRank estimates derived from overlapping 3-year rolling mobility windows. Prestige orderings were highly stable, with Spearman correlations exceeding ρ = 0.90 between adjacent windows and remaining above ρ ≈ 0.80 even across multi-year gaps. This result supports a cumulative-advantage system in which structural prestige differences among institutions are maintained over time, not merely driven by short-term hiring fluctuations.

## S7. Career-stage heterogeneity

To isolate mobility-related deviations from underlying career dynamics, we implement a two-stage correction.

- **Individual linear trend removal**: For each author *i*, we estimate a pre-move linear productivity trend using observations with event time $t<0$:

$$Y_{it} = a_i + b_i t + u_{it}$$

We subtract $b_i t$ from the outcome to remove individual career-stage growth.

- **Two-way fixed effects transformation**: The detrended outcome is further transformed via:

$$Y_{it}^* = Y_{it} - \bar{Y}_i - \bar{Y}_t + \bar{Y}$$

removing author and year fixed effects.

- **Event-study estimation**: Event-time coefficients are computed as:

$$\hat{\beta}_k = AVG(Y_{it}^* | EventTime = k)$$

with standard errors calculated as: $SE_k = \frac{SD(Y_{it}^*)}{\sqrt{N_k}}$, and 95% confidence intervals constructed using ±1.96·SE. All estimates are stratified by pre-move academic rank. Event time −1 serves as the reference period.

Replacing move direction with pre-move academic rank reveals clear career-stage heterogeneity in mobility effects (Fig. S2). Raw trajectories (top row) show that assistant professors are on steep upward productivity paths prior to mobility, whereas associates are comparatively stable and full professors exhibit flat or slightly declining trends. However, once we remove author fixed effects, year effects, and individual pre-move linear trends (bottom row), the apparent junior gains disappear. Assistant professors no longer exhibit post-move acceleration in publication output, citation impact, or top-cited output. The post-move coefficients for assistant professors are not statistically different from zero at any point in the event window.

Associate professors similarly do not have significant deviations from the baseline. All coefficients for productivity, citation impact, and the top 1% publication rate are close to zero and not significant in all three categories. For full professors, mobility is associated with a modest but persistent decline in publication volume (≈4–5%). However, citation impact rises steadily post-move, reaching roughly 25–35% above baseline five years post-move. Top-cited output also increases modestly, by approximately 1–2 percentage points over the same horizon.

## S8. Disciplinary heterogeneity

For each discipline *d*, we estimate:

$$Y_{it} = \sum_{k \neq -1} \beta_{kd} \cdot 1(EventTime_{it} = k) + \alpha_i + \gamma_t + \epsilon_{it}$$

This specification removes time-invariant productivity differences, macro publication growth, and promotion-related dynamics.

Although aggregate mobility effects on productivity are close to zero, discipline-specific dynamics reveal limited and temporally structured heterogeneity under the fully controlled specification (author fixed effects, year fixed effects, and time-varying rank controls). The strongest pre-move productivity decline occurs in Education three years prior to relocation ($\beta \approx -0.082$), corresponding to roughly an 8% reduction in output relative to the baseline year. Several other disciplines show modest negative coefficients in earlier pre-move years (e.g., Natural Sciences at event time −5; $\beta \approx -0.068$), but these effects attenuate once macro trends and rank progression are absorbed (Fig. S3).

On the post-move side, the largest positive effect appears in Applied Sciences four years after relocation ($\beta \approx 0.113$), implying approximately an 11% increase in publication output relative to baseline. Engineering displays a smaller but statistically significant gain five years post-move ($\beta \approx 0.057$), while Medicine & Health exhibits a comparable magnitude at the same horizon ($\beta \approx 0.057$).

In contrast, laboratory-intensive disciplines such as Natural Sciences do not exhibit sustained post-move productivity gains once year and rank effects are absorbed. The modest productivity differences observed in a few applied domains do not extend to research impact or breakthrough output. Under the same specification, log FWCI effects are uniformly small and statistically insignificant across disciplines, and no umbrella domain shows persistent increases in top 1% publication rates following mobility (Figs. S4 and S5).

These discipline-level findings reinforce the earlier move-type analysis, which demonstrated broadly neutral mobility effects across transitions. The evidence indicates that faculty mobility does not systematically enhance research performance. At most, mobility is associated with modest, field-specific adjustments in publication volume in a limited subset of applied disciplines, without corresponding improvements in impact or top-cited output.

## S9. Gender heterogeneity

We estimate a unified interaction event-study model:

$$Y_{it} = \sum_{k \neq -1} \beta_k \cdot 1(EventTime_{it} = k) + \sum_{k \neq -1} \delta_k \cdot [1(EventTime_{it} = k) \times Female_i] + \theta Rank_{it} + \alpha_i + \delta_t + \epsilon_{it}$$

Where:

- $\alpha_i$ is the author fixed effects (implemented via within transformation)
- $\delta_t$ is the year fixed effects
- $Rank_{it}$ is the time-varying academic rank
- Baseline $EventTime = -1$
- Standard errors clustered at the author level

The interaction coefficients $\delta_k$ capture female–male differences in mobility effects at each event time. This unified approach preserves comparability with the discipline-level model and avoids instability associated with sample splitting.

Identification relies on within-author variation over time. Author fixed effects absorb time-invariant productivity differences; year fixed effects absorb macro publication trends; rank controls account for career progression. The interaction terms isolate whether women experience systematically different dynamic responses to mobility relative to men.

When we restrict to the fully controlled event-study regression in the main analysis with author fixed effects, year fixed effects, and time-varying rank controls (Fig. S6), we find no gender-differential mobility effects for any of the outcomes (productivity, citation impact, or top-cited output) we investigated. The female–male gaps in log publication output remain constrained to be less than 3 percentage points in magnitude throughout the event window, and at none of the event times is the gap statistically different from zero. Similarly, the gender gap in log FWCI is constrained to stay within ±0.08 log points (≈8%) and exhibit no persistent directional pattern. Differences in top 1% publication rates remain within ±0.02 (≈2%) and show no sustained post-move divergence.

We find no evidence that mobility is associated with differential gains or losses for women in terms of publication volume, citation impact, or breakthrough output once career stage and macro trends are properly controlled. This finding is consistent with the aggregate, career-stage, and discipline-level results, reinforcing the conclusion that mobility is broadly performance-neutral across multiple performance dimensions.